\documentstyle[12pt]{article}
\begin{document}
\baselineskip .3in
\begin{titlepage}
\vskip .1in
\begin{center}{\large{\bf Fractal Space Time and Variation of Fine structure Constant  }}
\vskip .5in
 {\large{ A. Bhattacharya $^{\ddag}$,A. Chandra and B. Chakrabarti }}
\end{center}
\vskip .1in
\begin{center}
{\large{ Department of Physics, Jadavpur
University,Calcutta-700032, India. \\}}

\end{center}

\vskip .1in  The effect of fractal space time of the quantum
particles on the variation of the fine structure constant $\alpha$
has been studied. The variation of fine structure constant has
been investigated around De Broglie length $\lambda$ and compton
length $\lambda_{c}$ and it has been suggested that the variation
may be attributed  to the dimensional transition of the particle
trajectories between these two quantum domains. Considering the
Fractal universe with a small inhomogeneity in the mass
distribution in the early universe, the variation of the fine
structure constant have been investigated between matter and
radiation dominated era. The fine structure constant shows a
critical behaviour with critical exponent which is fractional and
shows a discontinuity. It has been suggested that the variation of
the fine structure constant may be attributed to the intrinsic
scale dependance of the fundamental constants of nature.

 \vskip .1in PACS: 98.80.Cq. 98.80.Bp. 98.65.Dx

 \vskip  .1in

$^{\ddag}$ E-mail pampa@jdvu.ac.in

\end{titlepage}

\newpage
{ The possibility of variation of fundamental constants of nature
has been a long standing problem and widely addressed  by a number
of authors [1] including Dirac himself [2]. It has been suggested
that the fine structure constant $\alpha$ =$e^{2}/c\hbar$ varies
with the age of the universe and the idea get a new impetus after
the discovery of CMB data and some cosmological observation of
data from supernova. With the progress of the observational
cosmology the experimental verification of the suggestions seems
to be plausible. Recent observations of the distant quasars have
suggested that the fine structure constant varies with
cosmological time scale and the variation is $\approx$
d$\alpha$/$\alpha$ $\approx$ $10^{-5}$ [4] over the time period
since the emission of the quasar light. Recent experimental
observation from Keck telescope has suggested that $\alpha$ varies
with the direction and it is not same all over the universe giving
an indication that the law of physics are not same everywhere [5].
The problem like horizon, flatness can be resolved if it can be
considered that speed of light was faster in the past. Huey et al
[6] studied CMB anisotropy spectrum dependence on fine structure
constant and equation of state (EOS) of the dark energy component
of the total energy of the universe. They have pointed out that
the varying $\alpha$ can be partially compensated by adjusting the
EOS of dark energy and obtained
($\alpha$-$\alpha_{0}$/$\alpha_{0}$) = $\triangle$ $\alpha$ /
$\alpha_{0}$ = 10$^{-2}$ to 10$^{-3}$ where $\alpha_{0}$ is the
current value. Barrow et al [7] have studied the dynamics of
varying $\alpha$ theories by introducing an exponential or inverse
power law for scalar field which allows the time variation of
$\alpha$. They have observed $\dot{\alpha}$ /$\alpha_{0}$  $\sim$
1.2.10$^{-16}$/yr. In a subsequent work [8]they have studied the
red shift dependence of $\alpha$ analyzing supernova results. They
have observed that during matter dominated epoch VSL mechanism
works where in the matter dominated era the varying  c effect
switched off allowing  $\Lambda$ to eventually surface resulting a
accelerating universe. They have found that the residual variation
in c induces a variation in $\alpha$ which agrees with the
observation. Anchordoqui et al [9] have investigated a
cosmological model which includes variation of $\alpha$ in short
time scale and observed a phase transition in neutrino mass at a
red shift z=0.5 which induces a phase transition in $\alpha$.
Moffat [10] has studied the cosmological evolution of $\alpha$
analyzing the absorption spectra of distant quasars considering a
cosmological scenario where the speed of light varies. Varying
$\alpha$ usually attributed to the varying speed of light whereas
varying e is a less radical approach. Bakenstein [11] studied a
theory which gives a varying e which preserves the local gauge and
Lorentz invariance and generally covariant. Berman et al [12]
investigated the time variation of $\alpha$ due to varying
electrical and magnetic permittivities. They have argued that the
present accelerating universe and exponential inflation may be
correlated with that variation. Sanvik et al [13] have
investigated varying $\alpha$ with varying e. They have observed
that $\alpha$ remains constant in radiation era, undergoes a small
change in matter dominated era and approaches a constant value
when universe starts accelerating due to positive cosmological
constant. Cingoz et al [14] have presented direct measurement of
temporal variation of $\alpha$ with atomic Disposium (Dy). The
result shows a fractional variation of (-2.7 $\pm$ 2.6)10$^{-15}$
without assumption of constancy of other fundamental constants.
Uzan [15] has made an review on varying $\alpha$ and indicated
that the variation may induce a new cosmological problem as
varying $\alpha$ cannot be naturally explained from field
theoretical approach.

 In the present work we have investigated the variation of
 $\alpha$ both in the context of the microphysics and cosmology. As we go below the
 classical radius of electron the nature of interaction changes
 and consequently variation in $\alpha$ is induced.
 We have studied the variation of $\alpha$ in the quantum domain
 near De Broglie length ($\lambda$) and the relativistic domain near the
 Compton length ($\lambda_{c}$) for electron.It has been
 suggested that the variation in $\alpha$ may be attributed to the
 fractal character of the space-time of trajectories of the quantum
 particles. The time variation of $\alpha$ between radiation dominated era and matter dominated era
  has also been investigated in the framework of the fractal universe studied by Banerjee et al [16] where small
 inhomogeneity has been incorporated through a mass fractal
 dimension (d) in the density distribution of FRW universe.
  It has been suggested that the scale dependence and power law
behaviour of the fundamental constants may
 be the manifestation of fractal character of the space time indicating the fact that
 geometry plays a fundamental role in describing an interaction.
 \newpage
 {\bf Variation of Fine structure constant in Quantum Domain--a
 Fractal Transition}:-
  The De Broglie wave length for an electron is given by:
  \begin{equation}
  \lambda= \frac{h}{mv}
  \end{equation}
The comptan wavelenght runs as:
  \begin{equation}
  \lambda_{c}=\frac{h}{2\pi mc}
  \end{equation}
  From (1) and (2) we get,
 \begin{equation}
  \lambda_{c}= \frac{\lambda}{2\pi} (\frac{v}{c})
  \end{equation}
  The first Bohr orbit can be expressed as:
\begin{equation}
  a_{0}=\frac{\lambda_{c}}{2\pi \alpha}
  \end{equation}

from (3) and (4) we obtain:
  \begin{equation}
  \alpha =\frac{\lambda}{4 \pi^{2} a_{0}} (\frac{v}{c})
  \end{equation}
  It is well known that the trajectory of a quantum particle is
  irregular in the fine scale and the path of the quantum mechanical particle
  is non differentiable. Feynman path integral approach [17] is a
  come back to the real situation of the quantum particle
  trajectories whereas in Bohr concept the trajectory has been
  completely abandoned.
  Feynman and Hibbs pointed out that with  $\epsilon$ =$\delta t$ is the resolution:
  \begin{equation}
  <\frac{x_{k+1}- x_{k}}{\epsilon}>^{2}= -\frac{\hbar}{im
  \epsilon}<l>
  \end{equation}
   The above expression (6) can be interpreted as that the transition
   element of the square of the velocity varies as 1/$\epsilon$
   and thus tends to infinity as $\epsilon$  $\longrightarrow$ 0 which is
   characteristic feature of a fractal measure. The expression (6)
   shows that the velocity of the particle is purely quantum
   relation not classical any more and $<v>^{2}$ tends to zero as
   $\epsilon$ tends to infinity [18].
    Now from the expression (6)we get:
    \begin{equation}
    \frac{dx}{dt}= <v^{2}>^{\frac{1}{2}}= (\delta t)^{-1/2}
     \end{equation}
     We can recast the above expression as:
      \begin{equation}
    \frac{dx}{dt}= = (\delta t)^{\frac{1}{D}-1}
     \end{equation}
     where D is the fractal dimension. According to the fractal definition the velocity is
     characterized by (dx/dt)$\propto$ ($\delta$x)$^{-1}$ [18].
     The derivative of the fractal function diverges like
     (($\delta$x)$^{D_{T}-D}$ which in turn can be represented as
     ($\delta$t)$^{\frac{1}{D}-1}$ depending on the choice of the
     variable which defined the resolution and $D_{T}$ is the
     topological dimension. From expression (7)
     the fractal dimension of a quantum trajectories is identified as D=2
     whereas
     the topological dimension is $D_{T}$=1 and as the particle undergoes transition from the classical
     domain to quantum domain the the transition
     may be explained as a consequence of a transition from non-fractal
     domain D=1 to fractal domain D=2.
 Now Equation (5)can be recast as :
 \begin{equation}
    \alpha= (\lambda/4 \pi^{2}a_{0}c)(\delta
    t)^{\frac{1}{D}-1}\sim (\delta t)^\frac{-1}{2}
     \end{equation}
     so that the first derivative of $\alpha$ varies as:
     \begin{equation}
    \dot{\alpha} = A'(\delta t)^{\frac{-3}{2}}
    \end{equation}

     Around the de Broglie length the fine structure constant varies as
     the $\delta t^{\frac{-1}{2}}$ as the particle undergoes a
     transition from classical to quantum domain. The time
     variation of $\alpha$ shows a fractal character with
     $\dot\alpha$ $\rightarrow$ 0 as the resolution $\delta$t
     tends to zero.

     In the non-relativistic domain around the compton wave length the
     momentum of the quantum relativistic path varies like p$\propto$ ($\delta$
     x)$^{-2}$ where $\delta x$ $\langle$ $\lambda_{c}$ [18] indicating a fast
     transition from quantum mechanical to quantum relativistic
     domain which leads to a fractal dimension 3 for the particle
     trajectory. With the similar argument as discussed above we
     get:
 \begin{equation}
    \alpha \sim (\delta x)^{D_{T}-D} \sim (\delta
    t)^{\frac{1}{D}-1} \sim (\delta t)^{\frac{1}{3}-1} \sim (\delta t)^\frac{-2}{3}
    \end{equation}
    so that,
    \begin{equation}
\dot{\alpha} \sim (\delta t)^\frac{-5}{3}
     \end{equation}
Showing fractal behavior as in the quantum domain. So it has been
observed that the transition from non relativistic to relativistic
domain  represent a transition from fractal space of dimension of
2 to a fractal dimension of 3. The critical exponent for the time
variation of $\alpha$ have been obtained as 3/2 and 5/3
respectively for two domains and shows a discontinuity. It is
interesting to note that the loss of notion in the position should
be described by new transition to different dimensional space. It
would be interesting to point out here that according to Feynman
prescription the trajectory back in time is interpreted as the
negative particles production. The dependence on the resolution or
the non differentiable path may be interpreted as virtual particle
creation and its contribution.

 \vskip  .1in

{\bf Variation of Fine structure constant in Fractal universe}:-

The FRW metric for homogeneous universe and Einstein Equation is
given by [19]:
\begin{equation}
\lbrack {\frac{\rm  dR(t)/dt}{\rm R}\rbrack }^2 = {\frac{\rm 4\pi
G \rho}{\rm 3}} - \frac{\rm K}{\rm R(t)^{2}} + \frac{\rm
\Lambda}{\rm 3}
\end{equation}
where the symbols have their usual meanings. Assuming $\Lambda$ to
be zero the above equation can be recast as:
\begin{equation}
\frac{\dot{R}^{2}}{2}-\frac{G \rho 4 \pi R^{2}}{3R}= -\frac{K}{2}
\end{equation}
Assuming K=0 for a Euclidean universe, the energy conservation
demands that:
\begin{equation}
\frac{d(\rho{R}^{3})}{dt}+ p \frac{d(R^{3})}{dt}= 0
\end{equation}
or
\begin{equation}
\frac{(\rho{R}^{3})}{dR}=-3pR^{3}
\end{equation}
where p is isotropic pressure.We have introduced a small
inhomogeneity in the mass distribution of the early universe for
large value of R where the energy density varies through $\rho(R)$
$\sim$ $R^{-d-3}$ where d is the mass fractal dimension and lies
between 0$<d<$1 for matter (d=0) and radiation dominated era (d=1)
respectively in our previous work [16] so that we have obtained;
\begin{equation}
 R(t) \sim t^{\beta}
\end{equation}
 where $\beta$ = 2/(d+3) and
\begin{equation}
 H = \beta t^{-1}
\end{equation}
 H is the Hubble factor with the conventional nomenclature of present time t as in
Weinberg [20] where t$_{0}$ $<$ H$_{0}$, H is the Hubble factor.
 In the context of the effective field theory and M theory, the change of the fine structure
  constant is obtained by coupling the dynamical scalar field $\phi$ to the photon kinetic term in the
  low energy effective action. Moreover to study the cosmology of the field $\phi$, it has been assumed
  that $\phi$ is governed by a lagrangian L = ($\delta$$\phi$)$^{2}$ - V($\phi$) where the potential energy
   is give by generic form $\mu$$^{4}$ $f$($\phi$/M) [21]. $\phi$ and M are microphysical parameters. Considering
    slow time variation of $\alpha$ the expression for d$\alpha^{-1}$ can be recast as [21]:

\begin{equation}
d\alpha^{-1} = \frac{\rm 4\pi\epsilon\mu^{4}}{\rm 3M^{2}}. \int
dtf^{'}(\phi/M)/ H
\end{equation}
With $\phi$/M approximately constant during matter dominated era
we may recast the above equation as:
\begin{equation}
d\alpha^{-1} = K/\beta \int dt/H
\end{equation}
 Now using the time dependance of Hubble parameter from the expression (18) obtained above we get,
\begin{equation}
 d \alpha^{-1} = \frac{K}{2\beta}t^{2}
\end{equation}
Integrating above expression we get,
\begin{equation}
 \alpha\sim \frac{2}{K(d+3)}t^{-3}
\end{equation}
 where $\beta$ = 2/d+3. The above expression shows the
variation of $\alpha$ in the mixed phase of matter and radiation
dominated era  in the limit 0 $<$ d $<$ 1.  We have observed that
$\alpha$ varies very sharp with time in the matter dominated
era.It is interesting to observe that $\alpha$ shows power law
behaviour with critical exponent 3.

{\bf Discussions and conclusions}:-
 In the present work we have
 investigated the variation of fine structure constant both in
 microphysics and in the early universe. We have tried to understand how the
 nature of the space time geometry particularly fractal space affects the variation $\alpha$ in
  the context of electrodynamics and different phases of early universe. Recently Culetto et al [22]studied the
   role of fractal
  geometry in scaling the fundamentals of electrodynamics. They have
  come across an expression for $\alpha$ which relates Feigenbaum's
  universal number and Thue-Morse constant.They have argued that
  the involvement of these constant points towards a digital
  regime at the infinitesimal level and one could not
  escape geometrization. They have also pointed out that the
  charge quantization may be tied to the fractal geometry. Nottale [18] has shown that
the fine structure in electrodynamics become logarithmically
divergent below compton length and this new behaviour has been
attributed to the electron positron pair creation and annihilation
which mainly occurs as expected from Heisenberg relation. Naschie
et al [23] have derived an expression for fine structure constant
from $\xi(\alpha)$ theory and pointed out that $\alpha$ can be
interpreted as quasi geometrical probability and
$\frac{1}{\alpha}$ is described as Housedroff dimension of
corresponding subspace. Goldfain et al [24] have derived fine
structure constant considering the fractional time evolution of
stochastic electrodynamics in the asymptotic limit of QED which is
defined as the large fluctuation limit of all relevant
variables.They have pointed out that the loss of characteristics
scale, typical anomalous diffusion and complex behaviour is linked
with the theory of critical phenimena.In their work fine structure
constant has been found to emerge from the fractional evolution of
density matrix whereas Chang et al [25] have proposed that the
Finsler space-time can account the fine structure constant
variation obtained from quasar spectra and suggested that the
variation can be attributed to the space time inhomogeneity and
anisotropy. the In the current work from the dimensional analysis
it has been found that the fine structure constant becomes
function of the resolution and obey a power law behaviour
diverging with the resolution in the quantum domain which may be
attributed to the creation of virtual pairs as the non
differentiability of the trajectory of the quantum particle
 means the new particle creation with the Feynman prescription. In
fractal space the trajectories are completely non differentiable
and there is no lower cut off below which fractralization would
stop. Hence there is always structure whatever the scale may be
and one never reaches the non structured limit that is assumed in
standard theory. This reopens the hope that the internal quantum
numbers are nothing but very internal structure of the particle
trajectory. However Cannatan et al [26] have pointed out that the
transition from quantum to relativistic quantum domain occurs from
D=2 to D=1 again not in D=3 which means again coming back to
non-fractal space time in the relativistic domain.

The variation of $\alpha$ in the early universe has been studied
  and it has been observed that the mass fractal dimension do not
  have much significant role. However it has been observed that the
  time variation of $\alpha$ obey an critical behaviour with
  critical exponent 3 in matter dominated era. BSBM [11] model predicts $\alpha$ $\sim$
  $t^{1/2}$ in radiation era and $logt$ in the matter dominated era
  predicting a constant value in accelerating era. Barrow et al [7-8]
   have made a detailed discussions on time variation of
  $\alpha$. Moffat [10] has obtained similar type of relation for
  $\Delta\alpha$ /$\alpha(0)$ and considered the exponent as 1.5.

    In the present work we have studied consequences of fractal space in the variation of fine structure constant
    and found that it may be termed as running coupling
    constant embedding the infinite distance limit
    characteristics. This may be the manifestation of the fact
    that the set has structure at every minute level and we can
    never end up without any structure however small the scale may be. The fractional charge may be also be related
    to the fractal path of the quantum particles [22]. It has been
    further suggested that the production of particle -anti
    particles may be the manifestation of intrinsic fractal behaviour of space in quantum domain.
 Starting from the cosmological scale to microphysics
    the variation of $\alpha$
   represents the scale relativity which is supposed to be the fundamental law of nature and the knowledge of
detailed space time structure is unavoidable in our understanding
of fundamental laws of Physics. However it
   may be pointed out here that the study of the variation of the fine structure constant needs the study of a
    domain which includes cosmology, astrophysics, high energy physics. Much more theoretical and observational
    efforts are needed to understand the origin of the variations of the fundamental constants of nature.

    Acknowledgement:
      This work is supported by University Grants Commission (UGC),
      New Delhi, India, grant no.F37-217/2009 (SR).

\pagebreak {\large\bf References:-}

\noindent [1]A. Albrecht  et al, Phys. Rev. {\bf 59}D,
043516(i999); Barrow J.D.et al, Phys. Lett.{\bf 443}B,104 (1998)
Phys. Rev. {\bf65}D, 063504 (2002).

\noindent [2]P.A.M.Dirac,  Nature {\bf 139}, 323 (1973).

\noindent [3] E.Teller, Phys. Rev.{\bf 73}, 801 (1948).

\noindent [4] J.K. Webb J.K. et al.,1999 Phys. Rev Lett {\bf 82},
884 (2001); 2001 Phys. Rev. Lett.{\bf 87}, 091301.

\noindent [5] M.T. Murphy et al, Mon. Not. Astron. Soc. {\bf
327}1208 (2001)

\noindent [6] G. Huey et al, Phys. Rev. {\bf 65}D 083001 (2003)

\noindent [7]J. D. Barrow et al. Astrophys. J . Lett.{\bf 532}L87
(2000).

\noindent [8] J.D. Barrow et al, Phys. Rev. {\bf 78}D, 083536
(2008)

\noindent [9] L. Anchordoqui et al, Phys. Letts. {\bf 660}B 529
(2008)

\noindent [10]J. W. Moffat, ArXiv: astro-ph/0109350v2

\noindent [11] J. D. Bakenstein, Phys. Rev. {\bf 25}D, 1527 (1982)

\noindent [12] M. S. Berman, Revista Mexicana de Astronomia y
Astrofisica {\bf 45} 139 (2009)

\noindent [13] H. B. Sanvik et al, Phys. Rev. Letts. {\bf 88},
031302(2002)

\noindent [14]C. Cingoz, Phys. Rev. Letts {\bf 98}, 040801 (2007)

\noindent [15] J. P. Uzan J.P, Rev. Mod. Phys. {\bf 75}, 403
(2003)

\noindent [16] S. N. Banerjee et al, Mod. Phys. Letts.A {\bf 12},
537 (1997)

\noindent [17]R. P. Feyman  and A.R. Hibbs, Quantum Mechanics and
Path Integrals( MacGraw-Hill),(1965).

\noindent [18] L. Nottle, Fractal spacetime and Microphysics,
World Scientific, Singapore,(1992) pp 95

\noindent [19] M.Ozer, Mod. Phys. Letts. {\bf 13}A, 571(1998)

\noindent [20] S. Weinberg, The Three Minutes (Harper Collin,
1993) p.180

\noindent [21] T.Bank et al, Phys. Rev. Letts. {\bf 88},131301
(2002)

\noindent [22] F.J. Culetto et al, 'Does Fractal geometry tune
electrodynamics' scales?' article in web.

\noindent [23] M.S. Naschie, Chaos, solitons and Fractals.{\bf
10}, 1947 (1999)

\noindent [24] E. Goldfain, Chaos,solitons and Fractal.{\bf 17},
811(2003)

\noindent [25] Zhe Chang et al, arXiv: 1106.2726

\noindent [26] F. Cannatan et al, Am. J. Phys {\bf 56}, 721(1988).

\end{document}